# A Lattice Evaluation of the Deep-Inelastic Structure Functions of the Nucleon[*]

M. Göckeler[1,2], R. Horsley[3], E.-M. Ilgenfritz[3], H. Oelrich[1], H. Perlt[4],
P. Rakow[5], G. Schierholz[6,1] and A. Schiller[4]

[1] Höchstleistungsrechenzentrum HLRZ, c/o Forschungszentrum Jülich,
D-52425 Jülich, Germany

[2] Institut für Theoretische Physik, RWTH Aachen,
D-52056 Aachen, Germany

[3] Institut für Physik, Humboldt-Universität,
D-10115 Berlin, Germany

[4] Fakultät für Physik und Geowissenschaften, Universität Leipzig,
D-04109 Leipzig, Germany

[5] Institut für Theoretische Physik, Freie Universität,
D-14195 Berlin, Germany

[6] Deutsches Elektronen-Synchrotron DESY,
D-22603 Hamburg, Germany

**Abstract**

The lower moments of the unpolarized and polarized deep-inelastic structure functions of the nucleon are calculated on the lattice. The calculation is done with Wilson fermions and for three values of the hopping parameter $\kappa$, so that we can perform the extrapolation to the chiral limit. Particular emphasis is put on the renormalization of lattice operators. The renormalization constants, which lead us from lattice to continuum operators, are computed perturbatively to one loop order as well as non-perturbatively.

---

[*]Talk given by G. Schierholz at the Workshop *QCD on Massively Parallel Computers*, Yamagata, March 1995

# 1 Introduction

For our theoretical understanding of the short-distance structure of the nucleon, as well as for a successful explanation of the recent HERA and polarized lepton-nucleon scattering data, a calculation of the deep-inelastic nucleon structure functions from first principles is needed. The theoretical basis for such a calculation is the operator product expansion. The quantities of primary interest are the lower moments of the quark and gluon distribution functions and their higher-twist counterparts.

We have initiated a program to compute the moments of the unpolarized, $F_1$ and $F_2$, and polarized nucleon structure functions, $g_1$ and $g_2$, on the lattice. First results of our calculation have been reported in Ref. [1]. In this talk we shall focus on two topics: the valence quark distribution and the spin content of the nucleon. The calculation of the gluon distribution functions and the distribution functions involving sea quarks is in progress. Because of space limitations we shall not be able to discuss the higher moments of $g_1$ and the structure function $g_2$.

Thus we shall concentrate on the moments $\langle x^{n-1} \rangle$, where $x$ is the fraction of the nucleon momentum that is carried by the quarks, and $\Delta q$, the quark spin contribution to the nucleon spin. We have

$$\langle x^{n-1} \rangle^{(f)} = v_n^{(f)}, \quad (1)$$

where

$$\frac{1}{2} \sum_{\vec{s}} \langle \vec{p}, \vec{s} | \mathcal{O}_{\{\mu_1 \cdots \mu_n\}}^{(f)} | \vec{p}, \vec{s} \rangle = 2 v_n^{(f)} [p_{\mu_1} \cdots p_{\mu_n} - \text{traces}],$$

$$\mathcal{O}_{\mu_1 \cdots \mu_n}^{(f)} = \left(\frac{i}{2}\right)^{n-1} \bar{q} \gamma_{\mu_1} \overleftrightarrow{D}_{\mu_2} \cdots \overleftrightarrow{D}_{\mu_n} q - \text{traces} \quad (2)$$

with $q = u(d)$ for $f = u(d)$. Here $\{\cdots\}$ means symmetrization of the indices. Furthermore we have

$$\Delta q = \frac{1}{2} a_0^{(f)}, \quad (3)$$

where

$$\langle \vec{p}, \vec{s} | \mathcal{O}_\sigma^{5(f)} | \vec{p}, \vec{s} \rangle = a_0^{(f)} s_\sigma,$$

$$\mathcal{O}_\sigma^{5(f)} = \bar{q} \gamma_\sigma \gamma_5 q. \quad (4)$$

The lattice calculation is performed on a $16^3 \times 32$ lattice at gauge coupling $\beta \equiv 6/g^2 = 6.0$. We work in the quenched approximation where one neglects the effect of virtual quark loops. We use Wilson fermions, and we compute simultaneously at three values of the hopping parameter, $\kappa = 0.155, 0.153$ and $0.1515$, so that we can extrapolate our results to the chiral limit. This translates into physical quark masses $m_q$ of roughly 70, 130 and 190 MeV, respectively. So far we have collected of the order 1000, 600 and 400 independent configurations at the three $\kappa$ values.

The lattice operators $\mathcal{O}$ are obtained from the operators in the euclidean continuum, up to factors of i, by replacing the covariant derivative in (2) by the lattice covariant derivative

$$\overrightarrow{D}_\mu(x,y) = \frac{1}{2}[U_\mu(x)\delta_{y,x+\hat{\mu}} - U_\mu^\dagger(x-\hat{\mu})\delta_{y,x-\hat{\mu}}]. \tag{5}$$

We first compute the two- and three-point correlation functions

$$\begin{align}
C_\Gamma(t,\vec{p}) &= \sum_{\alpha,\beta} \Gamma_{\beta,\alpha} \langle B_\alpha(t,\vec{p})\bar{B}_\beta(0,\vec{p})\rangle, \\
C_\Gamma(t,\tau,\vec{p},\mathcal{O}) &= \sum_{\alpha,\beta} \Gamma_{\beta,\alpha} \langle B_\alpha(t,\vec{p})\mathcal{O}(\tau)\bar{B}_\beta(0,\vec{p})\rangle. \tag{6}
\end{align}$$

As our basic proton operator we use (with $C = \gamma_4\gamma_2$ in our representation)

$$B_\alpha(t,\vec{p}) = \sum_{\vec{x},a,b,c} e^{-i\vec{p}\vec{x}}\epsilon_{abc}u_\alpha^a(x)(u^b(x)C\gamma_5 d^c(x)). \tag{7}$$

The nucleon matrix elements of interest are then obtained from the ratio

$$\begin{align}
R(t,\tau,\vec{p},\Gamma,\mathcal{O}) &= C_\Gamma(t,\tau,\vec{p},\mathcal{O})/C_{\frac{1}{2}(1+\gamma_4)}(t,\vec{p}) \\
&= \frac{1}{2\kappa}\frac{E_{\vec{p}}}{E_{\vec{p}}+m_N}\frac{1}{4}\mathrm{Tr}\,[\Gamma N \mathcal{J} N] \tag{8}
\end{align}$$

(for $0 \ll \tau \ll t$), where $N = (E_{\vec{p}}\gamma_4 - i\vec{p}\vec{\gamma} + m_N)/E_{\vec{p}}$ and $\mathcal{J}$ is defined by

$$\langle \vec{p},\vec{s}|\mathcal{O}|\vec{p},\vec{s}\rangle = \bar{u}(\vec{p},\vec{s})\mathcal{J}u(\vec{p},\vec{s}). \tag{9}$$

In order to increase the overlap with the ground state and to make the plateau region – by this we mean the region where the excited states have died out – in $\tau$ as broad as possible, we use 'Jacobi smearing' [2]. We smear both source and sink. In Fig. 1 we show the effective nucleon mass for $\kappa = 0.155$, i.e. our lightest quark mass, as given by $\ln(C(t)/C(t+1))$. We find a good plateau. Our results for the hadron masses are compiled in Table 1. The chiral limit is obtained by extrapolating in $1/\kappa$ to $m_\pi = 0$. If we assume that $m_\pi^2$ vanishes linearly with $1/\kappa \to 1/\kappa_c$, we obtain the critical value $\kappa_c = 0.15693(4)$.

|       | $\kappa$ | | |
|-------|----------|----------|----------|
|       | 0.1515   | 0.153    | 0.155    |
| $m_\pi$ | 0.504(2) | 0.422(2) | 0.297(2) |
| $m_\rho$ | 0.570(2) | 0.507(2) | 0.422(2) |
| $m_N$ | 0.900(5) | 0.798(5) | 0.658(5) |

Table 1: The hadron masses in lattice units at $\beta = 6.0$.

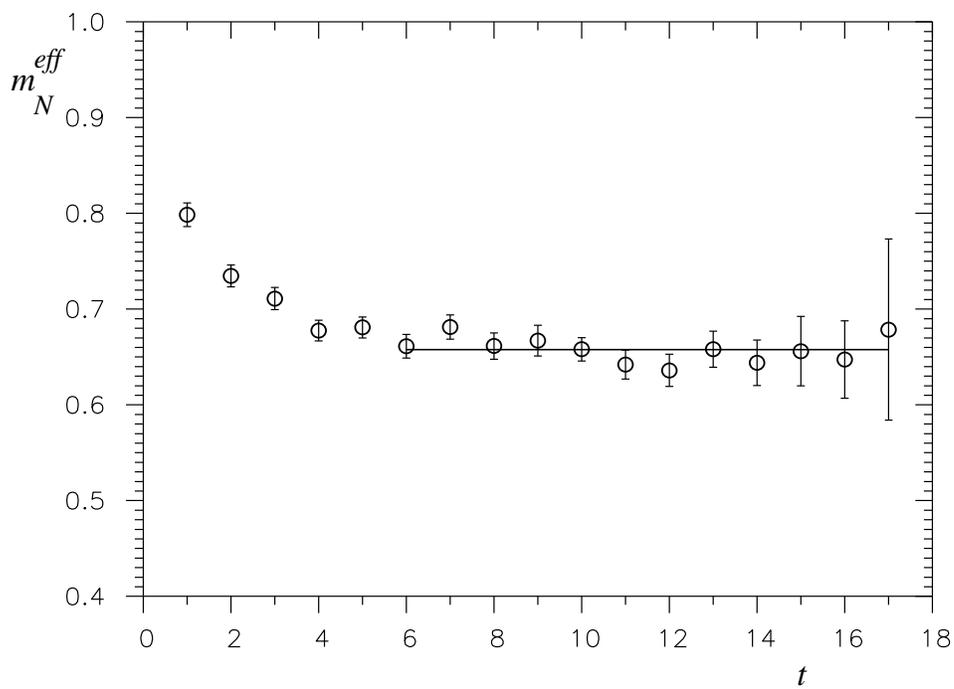

Figure 1: Effective nucleon mass plot. Both source and sink are smeared. The horizontal line indicates the result of the fit as well as the fit interval.

| Observable | $\langle\mathcal{O}\rangle$ | Components | $\gamma_{\mathcal{O}}$ | $B_{\mathcal{O}}$ |
|---|---|---|---|---|
| $\langle x \rangle$ | $v_2$ | $\mathcal{O}_{\{44\}} - \frac{1}{3}(\mathcal{O}_{\{11\}} + \mathcal{O}_{\{22\}} + \mathcal{O}_{\{33\}})$ | $\frac{16}{3}$ | -1.892(6) |
| $\langle x^2 \rangle$ | $v_3$ | $\mathcal{O}_{\{114\}} - \frac{1}{2}(\mathcal{O}_{\{224\}} + \mathcal{O}_{\{334\}})$ | $\frac{25}{3}$ | -19.572(10) |
| $\langle x^3 \rangle$ | $v_4$ | $\mathcal{O}_{\{1144\}} + \mathcal{O}_{\{2233\}} - \mathcal{O}_{\{1133\}} - \mathcal{O}_{\{2244\}}$ | $\frac{157}{15}$ | -37.16(30) |
| $\Delta q$ | $a_0$ | $\mathcal{O}_2^5$ | 0 | 15.795(3) |

Table 2: The lattice operators. The momentum is taken to be $\vec{p} = (2\pi/16, 0, 0) \equiv (p_1, 0, 0)$ in the case of $v_3$ and $v_4$ and $\vec{p} = 0$ elsewhere.

In our calculation of the three-point functions we have fixed $t$ at 13. For the unpolarized case we have taken $\Gamma = \frac{1}{2}(1 + \gamma_4)$. For the polarized case we have chosen $\Gamma = \frac{1}{2}(1 + \gamma_4)i\gamma_5\gamma_2$, which corresponds to polarization $+ - -$ in the 2-direction. For the calculation of the higher moments we need non-vanishing nucleon momenta. We have taken $\vec{p} = 0$ and $\vec{p} = (2\pi/16, 0, 0) \equiv (p_1, 0, 0)$.

## 2 Renormalization of Lattice Operators

The bare lattice operators, $\mathcal{O}(a)$, which are in general divergent, must be renormalized appropriately before we can use them. We define finite operators $\mathcal{O}(\mu)$, renormalized at the scale $\mu$, by

$$\mathcal{O}(\mu) = Z_{\mathcal{O}}((a\mu)^2, g(a))\mathcal{O}(a), \tag{10}$$

where we define

$$\langle q(p)|\mathcal{O}(\mu)|q(p)\rangle = \langle q(p)|\mathcal{O}(a)|q(p)\rangle \vert_{p^2=\mu^2}^{tree}, \tag{11}$$

$|q(p)\rangle$ being a quark state of momentum $p$. In the limit $a \to 0$ this definition amounts to the continuum, momentum subtraction renormalization scheme. The numbers that we will quote later on will all refer to this scheme.

The lattice operators must be constructed such that they belong to a definite irreducible representation of the hypercubic group $H(4)$ [3, 4]. In particular they must not mix with lower-dimensional operators. In this talk we will consider the operators listed in Table 2.

We have computed the renormalization constants of these operators (and others) in perturbation theory to one loop order [1, 5]. (See also Ref. [6].) We write

$$Z_{\mathcal{O}}((a\mu)^2, g) = 1 - \frac{g^2}{16\pi^2}C_F[\gamma_{\mathcal{O}} \ln(a\mu) + B_{\mathcal{O}}], \tag{12}$$

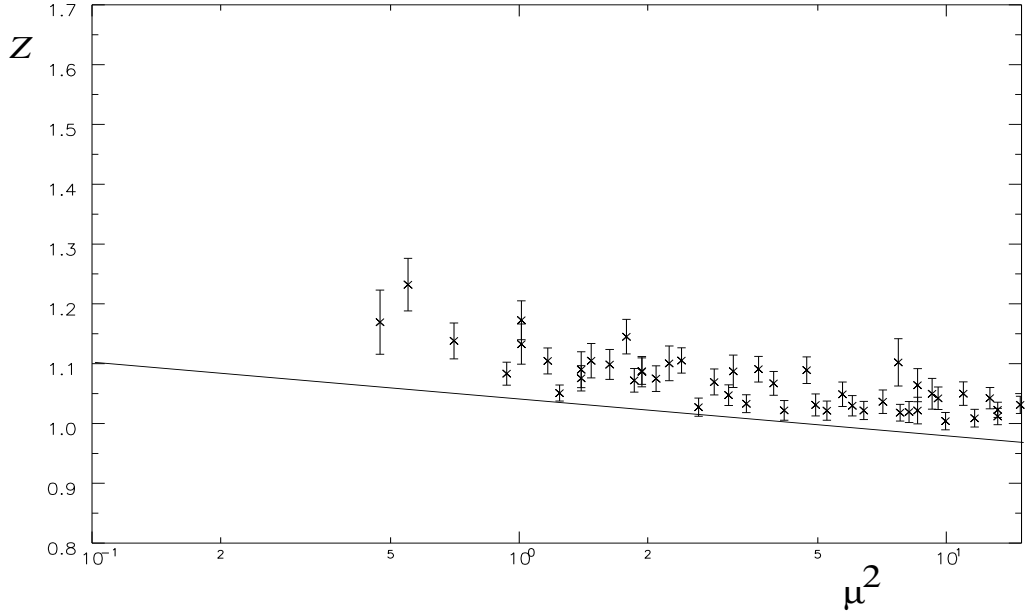

Figure 2: The non-perturbatively determined renormalization constant $Z_{v_2}$ as a function of $\mu^2$. The solid line represents the perturbative result given in Table 2.

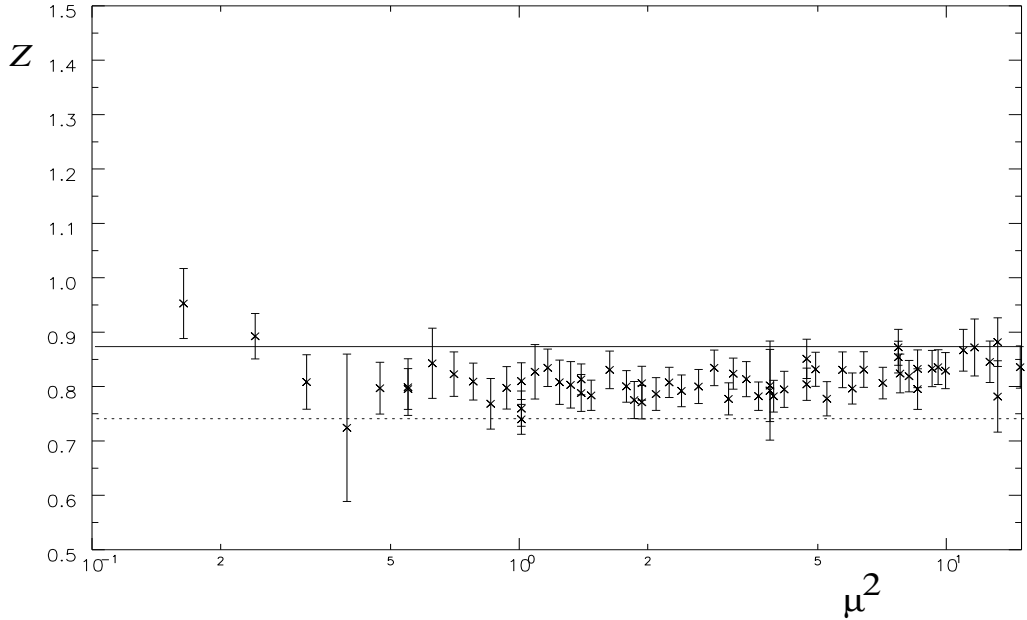

Figure 3: The non-perturbatively determined renormalization constant $Z_{a_0}$ as a function of $\mu^2$. The solid line represents the perturbative result given in Table 2. The dashed line is the result of tadpole improved perturbation theory.

where $C_F = 4/3$ and $\gamma_{\mathcal{O}}$ is the anomalous dimension that also enters the Wilson coefficients. The results of our calculation are listed in Table 2. In case of $v_3$ there is a small mixing problem [5], which we have ignored here, because numerically it is insignificant.

The effect of renormalization can be relatively large, in particular for those operators that involve higher powers of covariant derivatives. The main sources of contribution are the leg and operator tadpole diagrams. There has been a lot of discussion in the literature [7] on how to reorganize perturbation theory around these contributions in order to achieve a better convergence of the perturbative series.

It is important to determine the renormalization constants accurately. In view of this we have computed the renormalization constants non-perturbatively as well [8], following the suggestion of Ref. [9]. The renormalization constants are obtained from the calculation of the operator matrix element

$$\langle q(p)|\mathcal{O}(a)|q(p)\rangle \mid_{p^2=\mu^2}. \qquad (13)$$

For this calculation we need to fix the gauge. We have chosen the Landau gauge, and we average over all Gribov copies. In Figs. 2 and 3 we show our results for $Z_{v_2}$ and $Z_{a_0}$ as a function of $\mu^2$ (in lattice units). Here $\kappa = 0.153$, but we do not see any dependence on $\kappa$. We compare the non-perturbative results with the perturbative calculation. (The renormalization scheme used in the non-perturbative calculation differs slightly from the perturbative one. But the difference is insignificant [8].) We find remarkably good agreement between the two approaches. In Fig. 3 we have also shown the prediction of tadpole improved perturbation theory [7]. The non-perturbative result lies in between the perturbative and the tadpole improved result, so nothing seems to be gained by tadpole resummation in this case. But we will have to wait until we have computed all renormalization constants non-perturbatively before we can draw any firm conclusions.

In the following we shall take

$$\mu^2 = a^{-2} \approx 2\,\text{GeV}^2, \qquad (14)$$

which eliminates the logarithms in the renormalization constants, and we will denote $Z_{\mathcal{O}}(1, g = 1.0)$ by $Z_{\mathcal{O}}$.

## 3  A Selection of Results

We are now ready to compute the nucleon matrix elements (2) and (4). The physical matrix elements at the scale $\mu$ are obtained from the ratios (8) by

$$\begin{aligned}
R_{v_2} &= -\frac{1}{Z_{v_2}}\frac{1}{2\kappa}m_N v_2, & R_{v_3} &= -\frac{1}{Z_{v_3}}\frac{1}{2\kappa}p_1^2 v_3, \\
R_{v_4} &= \frac{1}{Z_{v_4}}\frac{1}{2\kappa}E_{p_1}p_1^2 v_4, & R_{a_0} &= \frac{i}{Z_{a_0}}\frac{1}{2\kappa}\frac{m_N}{2E_{p_1}}a_0.
\end{aligned} \qquad (15)$$

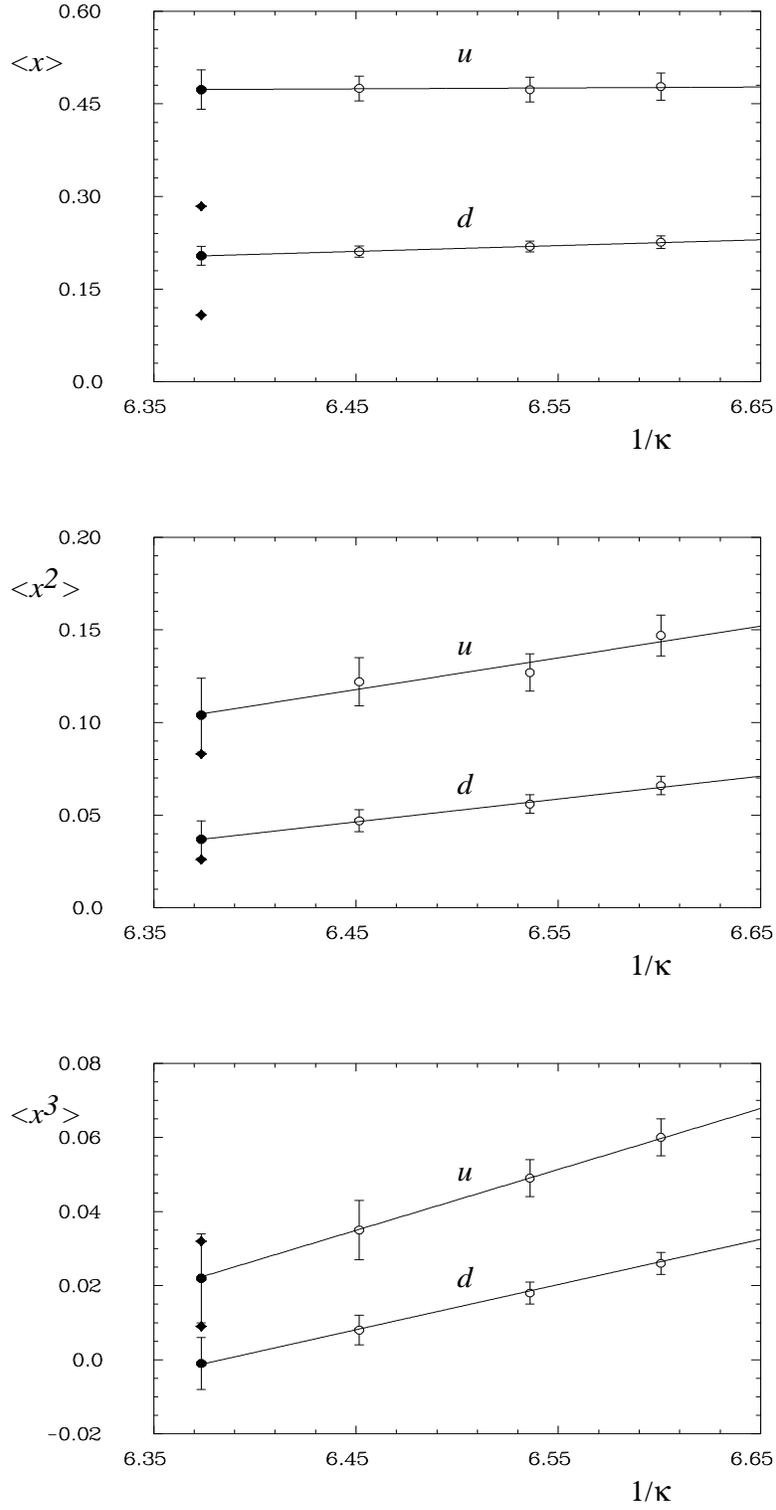

Figure 4: The moments $\langle x^{n-1} \rangle$ as a function of $1/\kappa$, together with a linear fit to the data. The solid circles indicate the extrapolation to the chiral limit. The diamonds mark the phenomenological valence quark distribution of Ref. [10] (fit $D_-$).

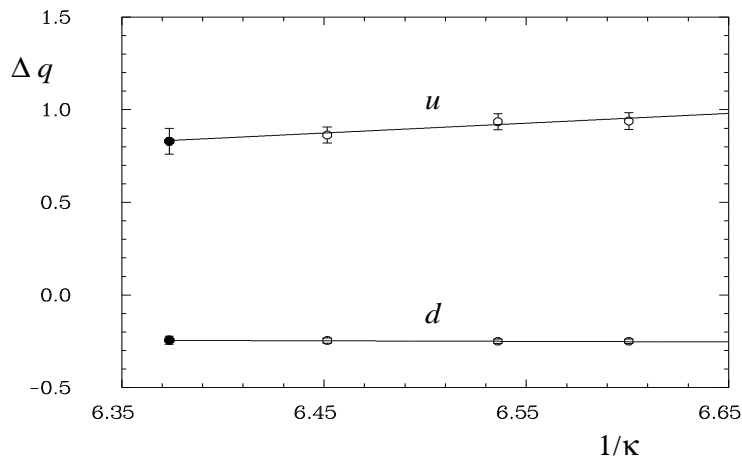

Figure 5: The quark spin contribution to the proton spin as a function of $1/\kappa$, together with a linear fit to the data. The solid circles indicate the extrapolation to the chiral limit.

We have made sure that we are computing the matrix elements of the lowest-lying state, i.e. the nucleon. We find that the signal is practically constant for distances of $\tau$ larger than two lattice spacings from the source ($t = 0$) and from the sink ($t = 13$). We have defined the continuum quark fields by $\sqrt{2\kappa}$ times the lattice quark fields. For the renormalization constants we take the perturbative values given in Table 2. The results are plotted in Figs. 4 and 5. All our results are given for the proton. The distribution functions of the neutron are obtained by interchanging $u$ and $d$.

The first observation we make is that the nucleon matrix elements show roughly a linear behavior in

$$1/\kappa \approx 1/\kappa_c + 1.8 m_q a, \qquad (16)$$

i.e. in the quark mass. The lines drawn in Figs. 4 and 5 are linear fits to the data. The result of the extrapolation is indicated by the solid circles.

Let us next discuss the moments $\langle x^{n-1} \rangle$. We find that the lowest moment is practically independent of the quark mass, while for growing $n$ the dependence on the quark mass increases. As the lower moments of the distribution functions are dominated by the small-$x$ region, this means that at small $x$ quark mass effects are negligible. At intermediate and large $x$, on the other hand, the distribution functions depend strongly on the magnitude of the quark mass. In the limit of large quark masses the moments approach the predictions of the non-relativistic quark model. In particular we find $\langle x^{n-1} \rangle^{(u)} \approx 2 \langle x^{n-1} \rangle^{(d)}$ for all $n$.

In Fig. 4 we have compared our results with the phenomenological valence quark distribution functions [10]. For the lowest moment the lattice result turns out to be significantly larger than the phenomenological value, while for the largest moment we find the opposite situation. This holds for both, $u$ and $d$ quark distributions. Thus the lattice calculation predicts a valence quark distribution that is larger at intermediate to small $x$ and smaller at large $x$ than the phenomenological distribution function. At present we have no explanation for this discrepancy.

Let us now discuss $\Delta q$. Sea quark effects may be neglected for heavy quarks, and they drop out in the difference $\Delta u - \Delta d$. In the chiral limit we obtain

$$\Delta u - \Delta d \equiv g_A = 1.07(9). \tag{17}$$

This is to be compared with the experimental value of the axial vector coupling constant $g_A = 1.26$. A recent lattice calculation of the sea quark contribution [12] finds $\Delta \bar{u} = \Delta \bar{d} = -0.14(5)$, $\Delta \bar{s} = -0.13(4)$, where we have used the perturbative renormalization factor. If we add these numbers to our results, we obtain for the total quark spin contribution to the nucleon spin

$$\Delta \Sigma = 0.18(8). \tag{18}$$

This is in agreement with the result of a full QCD calculation [13] which includes dynamical quark loops.

## 4 Conclusion

We have presented some results of a high statistics calculation of the lower moments of the polarized and unpolarized deep-inelastic structure functions of the nucleon. The calculation has been performed in the quenched approximation, and it was done for three different quark masses. This allowed us to extrapolate our results to the chiral limit.

The lattice data are rather accurate now, so that it is equally important to determine the renormalization constants precisely. We have computed the renormalization constants in perturbation theory to one loop order as well as non-perturbatively. So far we find consistent results.

The valence quark distributions that we have obtained on the lattice differ from the phenomenological distribution functions. This was also observed before [14]. One explanation could be that at smaller values of $Q^2$ higher twist contributions are non-negligible, which have not been included in the phenomenological analysis. We plan to investigate this possibility in the future. Our results for $\Delta q$ are consistent with experiment.

It is interesting to see how the results vary with the quark mass. At large quark masses our results agree largely with what one would expect on the basis of the quark model. For small quark masses there are, however, significant changes.

# Acknowledgments

This work was supported in part by the Deutsche Forschungsgemeinschaft. G. S. likes to thank A. Nakamura for his kind hospitality.

# References

[1] M. Göckeler, R. Horsley, E.-M. Ilgenfritz, H. Perlt, P. Rakow, G. Schierholz and A. Schiller, Nucl. Phys. **B** (Proc. Suppl.) **42** (1995) 337; DESY preprint DESY 95-128 (1995) (`hep-lat/9508004`).

[2] C. R. Allton et al., Phys. Rev. **D47** (1993) 5128.

[3] M. Baake, B. Gemünden and R. Oedingen, J. Math. Phys. **23** (1982) 944, *ibid.* **23** (1982) 2595 (E).

[4] J. Mandula, G. Zweig and J. Govaerts, Nucl. Phys. **B228** (1983) 109.

[5] M. Göckeler, R. Horsley, E.-M. Ilgenfritz, H. Perlt, P. Rakow, G. Schierholz and A. Schiller, in preparation.

[6] G. Martinelli and Y. C. Zhang, Phys. Lett. **B123** (1983) 433; S. Capitani and G. Rossi, Nucl. Phys. **B433** (1995) 351; G. Beccarini, M. Bianchi, S. Capitani and G. Rossi, Rome preprint ROM2F/95/10 (1995) (`hep-lat/9506021`).

[7] G. P. Lepage and P. B. Mackenzie, Phys. Rev. **D48** (1993) 2250.

[8] M. Göckeler, R. Horsley, E.-M. Ilgenfritz, H. Oelrich, H. Perlt, P. Rakow, G. Schierholz and A. Schiller, in preparation.

[9] G. Martinelli, C. Pittori, C.T. Sachrajda, M. Testa and A. Vladikas, Nucl. Phys. **B445** (1995) 81.

[10] A. D. Martin, W. J. Stirling and R. G. Roberts, Phys. Rev. **D47** (1993) 867.

[11] J. Ellis and M. Karliner, Phys. Lett. **B341** (1995) 397.

[12] M. Fukugita, Y. Kuramashi, M. Okawa and A. Ukawa, KEK Preprint 94-173 (1994); S. J. Dong, J.-F. Lagaë and K.-F. Liu, Kentucky Preprint UK/95-01 (1995).

[13] R. Altmeyer, M. Göckeler, R. Horsley, E. Laermann and G. Schierholz, Phys. Rev. **D49** (1994) R3087.

[14] G. Martinelli and C. T. Sachrajda, Nucl. Phys. **B316** (1989) 355; G. Martinelli, Nucl. Phys. **B** (Proc. Suppl.) **9** (1989) 134.